\documentclass[usegraphicx,usenatbib,useAMS]{mn2e}
\newcommand\lsim{\mathrel{\rlap{\lower4pt\hbox{\hskip1pt$\sim$}}
        \raise1pt\hbox{$<$}}}
\newcommand\gsim{\mathrel{\rlap{\lower4pt\hbox{\hskip1pt$\sim$}}
        \raise1pt\hbox{$>$}}}
\newcommand{\msun}{{\rm M_{\odot}}}

\usepackage{amsmath}
\usepackage{array}

\usepackage{graphicx}
\usepackage{fixltx2e}

\title[Critical Curve]{Beyond $J_{\rm crit}$: 
a critical curve for suppression of ${\rm H_2}$--cooling in protogalaxies}
\author[J. Wolcott-Green et al.]
{J. Wolcott-Green, Z. Haiman and G. L. Bryan\thanks{E-mail: jemma@astro.columbia.edu;
zoltan@astro.columbia.edu; gbryan@astro.columbia.edu}\\
Department of Astronomy, Columbia University, 550 West 120th Street, MC 5246, New York, NY 10027, USA}
\begin{document}

\date{}

\pagerange{\pageref{firstpage}--\pageref{lastpage}} \pubyear{2012}

\maketitle

\label{firstpage}

\begin{abstract}

Suppression of ${\rm H_2}$-–cooling in early protogalaxies has important 
implications for the formation of supermassive black holes seeds, the first 
generation of stars, and the epoch of reionization. This suppression can occur 
via photodissociation of ${\rm H_2}$ (by ultraviolet Lyman--Werner [LW] 
photons) or by photodetachment of ${\rm H^-}$, a precursor in ${\rm H_2}$
formation (by infrared [IR] photons). Previous studies have typically adopted 
idealised spectra, with a blackbody or a power--law shape, in modeling the 
chemistry of metal-–free protogalaxies, and utilised a single parameter, the 
critical UV flux, or $J_{\rm crit}$, to determine whether ${\rm H_2}$-–cooling 
is prevented. This can be misleading, and that independent 
of the spectral shape, there is a “critical curve” in the ($k_{\rm LW},k_{\rm H^-}$) 
plane, where $k_{\rm LW}$ and $k_{\rm H^-}$ are the ${\rm H_2}$–-dissocation rates 
by LW and IR photons, which determines whether a protogalaxy can cool below 
$\sim 1000$ Kelvin. We use a one–-zone model to follow the chemical and thermal 
evolution of gravitationally collapsing protogalactic gas, to compute this critical 
curve, and provide an accurate analytical fit for it. We improve on previous works 
by considering a variety 
of more realistic Pop III or Pop II-–type spectra from population synthesis models
and perform fully frequency–-dependent calculations of the ${\rm H_2}$-–photodissociation 
rates for each spectrum. We compute the ratio $k_{\rm LW}/k_{\rm H^-}$ for each spectrum, 
as well as the minimum stellar mass $M_\ast$, for various IMFs and metallicities, 
required to prevent cooling in a neighboring halo a distance d away. We provide 
critical $M_\ast/d^2$ values for suppression of ${\rm H_2}$-–cooling, with analytic 
fits, which can be used in future studies.
\end{abstract}


\section{Introduction}

It has long been known that the cooling of metal--free, primordial
gas, from which the first generation of stars form in the first
protogalaxies, is dominated by ${\rm H_2}$ molecules
\citep{Saslaw+Zipoy}.  Furthermore, the abundance of ${\rm H_2}$
molecules in an early protogalaxy is sensitive to radiation impinging
on the galaxy, and can be regulated by the early ultraviolet (UV)
background in the Lyman-Werner bands~\citep{HRL97}.  For unusually
soft spectra, infrared (IR) radiation can also play a role, through
the photo-detachment of the ${\rm H^-}$ ions, the main catalyst for
${\rm H_2}$ formation in primordial gas (e.g. \citealt{O01}).  Recent
simulations have suggested that the first stars may not be as massive
as had previously been thought (e.g. \citealt{Greif+11}).  In the
extreme case that the typical first-generation stars had masses as low
as a few ${\rm M_\odot}$, the IR radiation from these low-mass stars
would dominate radiative feedback and ${\rm H_2}$ chemistry in
protogalaxies in the early Universe \citep{WGH12}.\footnote{We do not
consider X-rays in this paper, which can also be important for early
${\rm H_2}$ chemistry \citep{HAR00,IT15}.}

In recent years, the suppression of ${\rm H_2}$ cooling via radiative
feedback has attracted a lot of attention, in the context of forming
massive black hole seeds in the early universe.  Observations of
high-redshift quasars reveal that supermassive black holes (SMBHs)
with masses of $\sim 10^9~\msun$ have already formed as early as
redshift $z\sim 7$ \citep{Mortlock+11}.  A promising way to form such
early massive SMBHs is to begin with a massive (say, $\sim 10^5\msun$)
seed BH at redshift $z\gsim 10$.  This massive seed can then grow
further by accretion, and reach $10^9~\msun$ by redshift $z\sim7$.  In
particular, in contrast to starting with a stellar-mass BH, the
accretion rate then need not exceed the value implied by the Eddington
limit (see recent reviews by
\citealt{VBreview12,SMBHreview13,NatarajanReview14} and \citealt{IHO15}).

A promising site for forming such massive BH seeds are in the nuclei
of so-called atomic-cooling halos -- i.e. dark matter halos with
virial temperatures $T_{\rm vir} \gsim 10^4{\rm K}$, or masses of
$\gsim {\rm few}\times 10^7 {\rm M_\odot}$.  Assuming that the gas at
the center of such a halo is metal-free (or with metalicity at most
$\sim 10^{-4}$ of the solar value; \citealt{OSH08}), and also that
${\rm H_2}$--cooling is disabled, the gas remains at temperatures near
$10^4{\rm K}$.  This leads to accretion rates in the core of the halo
as high as $\sim 0.1-1 {\rm M_\odot yr^{-1}}$. Several works have
argued that under these conditions, fragmentation and Population III
star-formation may be avoided, and a massive seed BH is produced
instead, either by direct collapse, or via an intermediate state of a
supermassive star
\citep{OH02,BL03,KBD04,LN06,SS06,BVR06,VLN08,WA08,RH09b,SSG10,SBH10,WGH11,WHB11,Latif+14}.\footnote{Recent
  simulations \citep{Regan+14} have suggested that fragmentation might
  occur at spatial resolution higher than currently numerically
  feasible.  On the other hand, even if fragmentation occurs, a
  supermassive star may still be the natural outcome, as long as ${\rm
    H_2}$ cooling is disabled, owing to the rapid migration and
  coalescence of the central fragments \citep{IH14,Hosokawa+15}.}

A crucial assumption in these scenarios is the lack of ${\rm
  H_2}$--cooling.  In the presence of ${\rm H_2}$, the gas in the
atomic cooling halos would likely fragment and form Population III
stars, similar to the case of lower-mass ``minihalos''
\citep{ABN02,BCL02,Yoshida+03}.  It has been well-established in both
semi--analytic studies and three--dimensional simulations that a
sufficiently strong dissociating LW flux can suppress ${\rm H_2}$--
cooling entirely, keeping the gas close to the virial temperature of
the halo throughout the initial stages of collapse
\citep{O01,BL03,RH09a,SSG10,SBH10}.  As demonstrated in these papers,
the spectral shape of the incident radiation is important in
determining whether this scenario is plausible.

In nearly all previous studies, the incident radiation field is
modeled either as a power--law, or as a blackbody spectrum with
temperature $10^4 \leq {\rm T_*/K} \leq 10^5$, and a critical flux
$J_{\rm crit}$ is defined as the minimum intensity required to prevent
cooling. Quoted at the Lyman--limit, $J_{\rm crit}$ for a $T_* =
10^5$K blackbody (hereafter referred to as T5) is usually found to be
in the range $10^3- 10^4$ in the customary $J_{21}$ units $J{\rm
  (13.6eV)} = J_{21} \times 10^{-21} {\rm erg s^{-1} Hz^{-1} cm^{-2}
  cm^{-2} sr^{-1}}$.

For the softest blackbody spectra considered, with $T_*=10^4$K
(hereafter T4), the nominal critical flux is much lower than for the
T5 type, typically $J_{\rm crit}\sim30$ (e.g. \citealt{O01}).  In the
T4 case, photodetachment of ${\rm H^-}$, a precursor in the primary
formation reaction for ${\rm H_2}$, causes suppression of ${\rm
  H_2}$--cooling, rather than photodissociation. This is due to the
large IR flux near the photodetachment threshold (${\rm h \nu \approx
  1-2}$eV) in a T4 spectrum compared to the T5 spectrum {\it for a
  fixed $J_{21}$}. However, as shown by \citet{WGH12}, this can be
misleading: the mass in stars must, in fact, be higher for a T4--type
population to produce the same $J_{21}$ as a T5--type.  Thus, the
lower value of $J_{\rm crit}$ for the soft T4 spectrum makes it more
difficult, rather than easier, to suppress ${\rm H_2}$--cooling.

The main goal of this paper is to investigate the ${\rm H_2}$--cooling
in the case of a ``realistic'' Pop II spectrum for the incident
radiation field, rather than an idealised spectrum such as a blackbody
or power--law. A single flux, $J_{\rm crit}$, is sufficient for
determining the cooling history for a blackbody (power--law), because
for a given blackbody temperature (exponent), there is a fixed ratio
of flux in the LW bands to the flux at the photodetachment threshold
($0.76$eV).  However, for more realistic spectra from population
synthesis modeling, it is necessary to consider separately the ${\rm
  H_2}$--photodissociation ($k_{\rm LW}$) and ${\rm
  H^-}$--photodetachment rates ($k_{\rm H^-}$), and their evolution
over time.

We show in \S~\ref{Sec:Results} that {\it there is, in general, a
  ``critical curve,'' rather than a single $J_{\rm crit}$}, for
determining whether ${\rm H_2}$--cooling is suppressed and thus
whether the halo is a candidate for a DCBH.  This critical curve is a
line in the $k_{\rm LW}$ vs $k_{\rm H^-}$ plane, and is independent of
the spectral shape.  We consider a range of population synthesis model
spectra from {\sc starburst99} and show how the results compare, with
respect to the critical curve, with results from blackbodies ($T_* =
10^4-10^5$K). We also provide an updated criterion for determining
whether ${\rm H_2}$--cooling is suppressed, as a fitting formula for
the critical curve.

The non--existence of a single $J_{\rm crit}$ and its importance for
time--evolving spectra from population synthesis models has also been
pointed out recently by \citet{AK15} and \citet{Agarwal+16}. These
studies, as well as \citet{Sugimura+14} have investigated the relative
importance of $k_{\rm LW}$ and $k_{\rm H^-}$ from {\sc starburst99}
and from blackbody spectra in suppressing ${\rm H_2}$--cooling.  We
here improve on these studies by computing the ${\rm
  H_2}$--photodissociation rate, summing over all relevant LW lines,
rather than assuming that the spectrum is flat in the LW bands, or
taking an average of the LW flux. We show that both previous
approximations, though computationally efficient, introduce
significant errors in the photodissociation rate compared to our full
frequency--dependent calculation.

This paper is organised as follows: In \S~\ref{Sec:Model} we describe
the details of our numerical modeling, We present our results using a
variety of {\sc starburst99} models in \S~\ref{Sec:Results}, along with 
an updated fitting formulae for both the critical curve and critical 
mass in stars.  We briefly discuss the results and implications of our 
work for future studies, and offer our conclusions in \S~\ref{Sec:Conclusions}.

\section{Modeling}
\label{Sec:Model} 

\subsection{One-Zone Model}

We use a one--zone model to follow the thermochemical history 
of primordial gas in a collapsing atomic cooling halo. This model 
has been shown to mimic the evolution found in three--dimensional 
hydrodynamical simulations very well \citep[][hereafter SBH10]{SBH10}
and is described in detail in, e.g. \citet{OSH08}. We briefly 
summarize the model here and highlight a few important updates 
we made to the chemistry network.  

The one--zone model prescribes a homologous spherical collapse, 
in which the dark matter evolution is that of a top hat 
overdensity with turnaround redshift $z_{\rm ta} = 17$ and
$\rho_{\rm DM}$ constant after virialization. The timescale of
collapse is taken to be the free--fall time. Using the solver 
{\sc lsodar}, we follow the thermal and chemical evolution of 
initially primordial gas, with nine chemical species included:
${\rm H, H^+, He, He^+, He^{++}, H^{-}, H_2^+, H_2}$, and electrons. 
Deuterium species are not included, as they do not affect our 
results at the temperatures of interest here, which are too high 
for HD--cooling to be important. Compressional heating as well 
as photochemical heating and cooling processes are included as 
detailed in SBH10. 

We have made several updates to the one--zone model, most of which have 
not been included in previous studies \citep[][is an exception]{Agarwal+16}. 
We use the modified self--shielding expression provided by \citet{WHB11}, 
which is more accurate for gas near the critical density, ${\rm n \gsim 10^3
  cm^{-3}}$.  The impact of a non--LTE ro-vibrational distribution is
discussed in \S~\ref{Sec:Caveats}.  \citet{WHB11} showed that the
column density approximation most often used in the one--zone model
for the optically--thick $k_{\rm LW}$ overestimates shielding by about
an order of magnitude in the density range of interest compared to 3D
simulations.  We therefore modify the column density as recommended in
that study: ${\rm N_{H2}} = 0.25 \times \lambda_{\rm Jeans} \times
  n_{\rm H2}$, which is half of the commonly used value.  Here
$\lambda_{\rm Jeans}$ is the Jeans length and ${\rm n}_{\rm H2}$ is the
number density of molecular hydrogen. Though this is a somewhat crude
fix, it more closely matches the results for optically--thick gas
found in 3D hydrodynamical simulations. One caveat should
be noted here: in general, ${\rm H_2}$ self--shielding behavior will 
depend on the shape of the incident spectrum, since the shape 
determines the relative weights of the LW transitions. However, in 
practice, we find that the self--shielding function for a given 
SB99 spectrum differs very modestly from that for a flat spectrum, 
with errors typically of order a few per cent for young bursts, 
and not larger than $\lsim 25$ per cent for the spectra considered here. 

We have also updated some of the most important chemical rates which
determine the ${\rm H_2}$ abundance, based on the results of recent
studies. These include the associative detachment rate (${\rm H + H−
  \rightarrow H_2 + e^−}$; \citealt{Kreckel+10}), and the mutual
neutralization rate (${\rm H^- + H^+ \rightarrow 2H}$;
\citealt{Stenrup+09}). We have added several reactions to the
one--zone chemistry network which have been shown to be important by,
e.g.  \citet{Glover15a,Glover15b}. These include collisional
dissociation of ${\rm H_2}$ by neutral hydrogen through dissociative
tunneling, the rate for which is provided by \citet{MSM96} and the
ionization of atomic hydrogen by H-H collisions and H-He collisions,
using the rates from \citet{Lenzuni+91}. The radiative recombination
rate for ${\rm H^+}$ is the Case B rate from \citet{Hui+Gnedin} --
appropriate for low--ionization fraction of gas collapsing in an
atomic cooling halo \citep{Glover15b} -- rather than the Case A rate
previously used in SBH10.  We use the rate for radiative association
from \citet{Abel+97}, rather than that from \citet{Hutchins76} used by
SBH10. The remainder of the chemistry network is the same as in SBH10
and provided in their Appendix.

\subsection{Model Spectra}

We use the publicly available population synthesis package 
{\sc starburst99} \citep[][hereafter SB99]{Leitherer+99} 
to generate spectra with a range of metallicities, IMFs, and ages.
We use the UV line spectra provided by SB99 (at 0.3 ${\rm \AA}$ 
resolution) for calculating the ${\rm H_2}$--photodissociation rate. 
Unfortunately, high--resolution spectra are not available for 
the full wavelength range required for all photochemical rates; 
therefore, for all continuum processes we use the SB99 spectra 
at lower resolution, which varies from (1-20 ${\rm \AA}$) over 
the relevant wavelengths. 

We focus here on spectra for the SB99 ``burst'' models, in which a
single population with specified mass, metallicity and IMF forms
instantaneously and evolves over time, without any ongoing star
formation. In the DCBH scenario, these are likely the most relevant,
because the intense flux required for ${\rm H_2}$--cooling suppression
must come from a very bright close neighbor, rather than from the
cosmological background (see e.g. 
\citealt{SBH10, Dijkstra+08, Ahn+09,Agarwal+12, Agarwal+14,Regan+14b,Habouzit+16}). 
These studies show that the most likely candidates for DCBHs are within 
1-2kpc of a bright source.  As pointed out by \citet{VHB14b}, one 
promising scenario is that two atomic cooling (sub-)halos in very close
proximity are nearly synchronised in their collapse, so that the first
to form stars illuminates the other with a strong LW flux mostly from
young, bright, massive stars. \citet{VHB14b} and \citet{Regan+16} show that there are
issues of photoevaporation if the collapsing DCBH candidate halo has
been irradiated with a strong flux for $\gsim 20$ Myr; we therefore
focus primarily on burst populations with ages up to 20 Myr.

\begin{table}
  \begin{center}
    \caption{Summary of {\sc starburst99} model galaxy parameters. 
    All assume a power-law IMF with exponent $\alpha$ and mass limits 
    given below in units of ${\rm M_\odot}$. Each IMF is run with
    metallicities Z = 0.001, 0.004, 0.008 ${\rm Z_\odot}$.}
    \label{tbl:SB99Models}
    \begin{tabular*}{0.475\textwidth}{@{\extracolsep{\fill}} l l l l l l }
      \hline
      Case &   A    & B    & C    & D & E \\
      \hline
      Limits & (1,120) & (10,120) & (30,120) & (30,500) & (100,500) \\
      $\alpha$ & 2.35 & 2.35 & 2.35 & 1 & 1 \\
      \hline\\
    \end{tabular*}
  \end{center}
\end{table}
We summarize the IMF parameter choices for our SB99 models in Table
\ref{tbl:SB99Models}. Each IMF is run at three different metallicities
from ${\rm Z/Z_\odot = 0.05-0.4}$, though for the direct collapse
model, the lower end of the range is more likely to be relevant, as
the DCBH halo must be nearly pristine (see, e.g. \citealt{OSH08}), 
and is less likely to remain so with a highly enriched near neighbor.

The specific luminosities at the Lyman limit ($L_{13.6}$ in units 
${\rm erg~s^{-1}~Hz^{-1}}~{\rm M_\odot}^{-1}$) output by SB99 are 
converted to $J_{21}$ by:
\begin{equation}
J_{21} = \frac{M_*}{16\pi^2d^2}\times 
\frac{L_{13.6}}{10^{-21}}, 
\end{equation}

where $M_*$ is the total stellar mass of the SB99 burst galaxy and $d$
is the distance from the illuminating halo.

\begin{figure}
  \includegraphics[clip=true,trim=0.in 0.1in 0.9in 1.in,
    height=2.8in,width=3.2in]{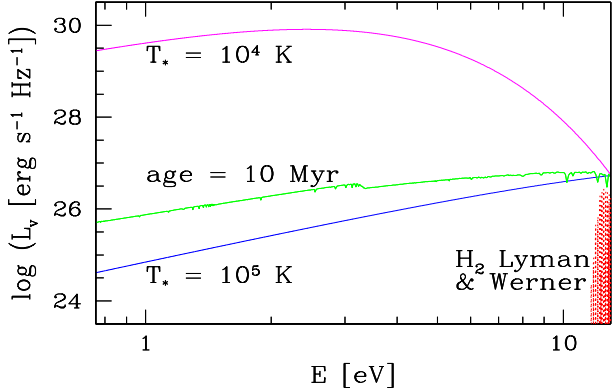}
  \caption{The spectrum of a {\sc starburst99} instantaneous burst
    model with stellar mass $M_*= 10^6 {\rm M_\odot}$, IMF A, metallicity
    $Z/{\rm Z_{\odot}} = 0.05$, and age $t=10$ Myr is shown in
    green. Spectra of blackbodies with ${\rm T_* = 10^4}$ (T4) and
    $10^5$ K (T5) are shown, scaled to the same luminosity at the
    Lyman limit. Red (dotted) vertical lines show the Lyman-Werner
    band transitions (arbitrary units). We show the spectra in the
    energy range of interest, from the ${\rm H^-}$ photodetachment
    threshold, 0.76 eV, to the Lyman limit, 13.6eV. The SB99 spectrum
    is much closer to T5 than to T4.}
  \label{Fig:Spectra}
\end{figure}

In Figure \ref{Fig:Spectra}, we show an example SB99 spectrum with
metallicity ${\rm Z/Z_\odot = 0.05}$ at ${\rm t= 10}$Myr, as well as
T4 and T5 blackbodies normalised to have the same luminosity at the
Lyman limit. The frequencies of the ${\rm H_2}$ LW transitions are
also shown for reference. As this figure illustrates, the SB99
spectrum is much closer to the T5 spectrum at the energies most
important for ${\rm H^-}$--photodetachment, 1-2eV. In
\S~\ref{Sec:Results} we discuss further the comparisons between SB99
spectra and blackbodies and the implications for ${\rm H_2}$--cooling.

\subsection{${\bf H_2}$--Photodissociation Rate} 

We calculate the frequency--dependent optically--thin
${\rm H_2}$--photodissociation rate for each of the SB99 models 
using the high--resolution UV spectra. Photodissociation occurs 
via the two--step Solomon process \citep[Solomon 1965; see also][]{FSD66,SW67}, 
in which ${\rm H_2}$ molecules are electronically excited by 
absorption of Lyman--Werner photons (900-1100 ${\rm \AA}$). 
Following excitation, $\sim 15\%$ of molecules are dissociated 
in subsequent radiative decay to the vibrational continuum of 
the electronic ground state, while all others return to a bound 
state via radiative cascade.

The rate of photodissociation of molecules from initial 
ro-vibrational state ({\it v,J}) is   

\begin{equation}
k_{{\rm diss},v,J} = \sum_{\it v',J'} \zeta_{\it v,J,v',J'}
{\it f}_ {{\rm diss},v',J'},
\end{equation}
where ${\it f}_{\rm diss,v',J'}$ is the fraction of molecules
that are dissociated from the excited state ({\it v',J'}), provided
by \cite{ARD00}. The ``pumping rate'' of ${\rm H_2}$ from 
the ground state ({\it v,J}) to ({\it v',J'}) is: 

\begin{equation}
\zeta_ {\it{v,J,v',J'}} = \int_{\it{\nu_{\rm min}}}^{\nu_{\rm max}}4\pi \sigma_{\it v,J,v',J'}(\nu)\frac{J_\nu}{h\nu}{\rm d}\nu;
\end{equation}
where $J_\nu$ is the specific intensity of the SB99 spectra in
the usual units and $h$ is Planck's constant. The limits of 
integration are from $h\nu_{\rm min}=11.1$eV to  $h\nu_{\rm max}=13.6$eV.
We use the data from \citet{ARL93a} to obtain the frequency--dependent
Lyman--Werner absorption cross--sections $\sigma_{\it v,J,v',J'}(\nu)$.

The total ${\rm H_2}$--photodissociation rate depends on the
ro-vibrational distribution of molecules, ${\it f}_ {v,J}$:
\begin{equation}
k_{\rm LW} = \sum_{\it v,J} k_{{\rm diss},v,J} {\it f}_ {v,J},
\end{equation}
which in general can depend on the detailed thermochemical history of
the collapsing cloud. For simplicity, we assume all molecules are in
the ground vibrational state with a Boltzmann distribution over
rotational levels (up to {\it J}=29). Although the critical densities
for rotational levels $J \gsim 5$ are larger than in our models 
(${\rm n_H \gsim 10^4 cm^{-3}}$), the populations in those levels are
vanishingly small for a Boltzmann distribution at $\sim 10^3$K.  We
further discuss the implications of a non--LTE ro-vibrational
distribution in \S~\ref{Sec:Caveats}.

\subsection{Rate of ${\bf H^-}$ Photodetachment}
The rate of ${\rm H^-}$ photodetachment is computed with the widely-used
\citet{SK87} fit for the cross-section data from \citet{Wishart79}. 
This is convolved with the flux from a blackbody or SB99 spectrum: 
\begin{equation}
k_{\rm H^-} = 4 \pi n_{\rm H^-} \int_{\rm 0.76~eV}^{13.6 eV} 
\sigma_{\nu,{\rm H^-}} \frac{J_\nu}{h\nu}{\rm d}\nu.
\end{equation}
\citet{Miyake+10} pointed out that a series of auto--detaching resonances
above $\sim 11$eV also contribute to the photodetachment rate; however, 
they find the additional suppression of the ${\rm H_2}$ abundance is only 
$\lsim 20$ per cent for blackbody spectra with ${\rm T_{BB}}$ up to $10^5$K, 
and we therefore do not include these resonant contributions to the rate. 

\section{Results and Discussion}
\label{Sec:Results}

\subsection{Approximate Photodissociation Rates}
\label{Sec:ApproxRates}

Due to the computational expense of calculating the
frequency--dependent ${\rm H_2}$ photodissociation rate as described
above, nearly all previous studies have instead derived an approximate
rate by assuming that the (incident) flux across the LW bands is flat. 
With the flux at 13.6eV specified by $J_{21}$ in the usual units, the
optically--thin rate is then parameterised as
\begin{equation}
k_{\rm LW} = 1.39 \times 10^{-12} \beta J_{21}.
\label{Eq:DefineBeta} 
\end{equation}
By convention, the rate is calculated assuming a flat spectrum in 
the Lyman--Werner bands and $\beta$ is usually defined as the ratio 
of the flux at a single LW frequency to the flux at the Lyman limit,
$\beta = J_{\rm LW}/J_{21}$. The most common choice is to define
$J_{\rm LW}$ as the flux at 12.4 eV, the midpoint of the Lyman--Werner
bands; we refer to this as $\beta_{\rm flat}$
\citep[e.g.][]{SBH10}. Alternatively, some studies define $J_{\rm LW}$
as the average flux in the LW bands \citep[e.g.][]{AK15}; we will
refer to this as ${\rm \beta_{\rm avg}}$:
\begin{equation}
\beta_{\rm avg} = \frac{1}{\Delta \nu_{\rm LW}} 
\int_{\nu_{11.1eV}}^{\nu_{13.6eV}}J_\nu {\rm d}\nu
\end{equation}

These approximations save significant computational expense, avoiding
integration over dozens or hundreds of LW lines, as described in the
full rate calculation above. However, even in the idealised case of a
$10^4$K blackbody spectrum, a non--zero slope of the spectrum in the
LW bands can lead to errors in $k_{\rm LW}$ of a factor of two
compared to the rate if the spectrum is assumed flat at 12.4
eV. Hotter blackbody spectra are flatter in the LW bands so the errors
are smaller, and only a few percent for ${\rm T_{BB} = 10^5}$K.  For
more realistic spectra, however, such as those from population
synthesis codes like SB99, both the non--zero slope as well as
absorption features -- e.g. O {\sc iv} $\lambda1035$, Ly$\beta$, C
{\sc ii} $\lambda$1036, and LW lines themselves \citep{Leitherer+99}
-- near the LW transition frequencies can cause the rate derived from
$\beta_{\rm flat}$ or $\beta_{\rm avg}$ to deviate significantly from
frequency dependent (``true'') rate.

\begin{figure}
  \includegraphics[clip=true,trim=0.in 0.1in 0.3in 1.in,
    height=2.8in,width=3.2in]{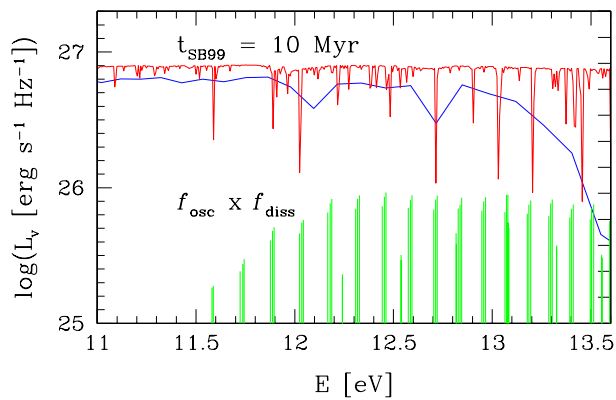}
  \caption{High- (red) and low-resolution (blue) spectra in the
    Lyman--Werner bands for a {\sc starburst99} model of an
    instantaneous burst with IMF A and ${\rm Z/Z_\odot} = 0.05$ at
    10 Myr. The Lyman--Werner transition strengths are shown in
    green. Because the absorption lines in the high--resolution
    spectrum overlap with the LW transition energies, significant
    errors can result from the assumption that the spectrum is flat in
    the LW bands, even when, as is the case here, the stellar
    population is relatively young and total absorption is modest.}
  \label{Fig:Spectra2}
\end{figure}

In Figure \ref{Fig:Spectra2} we show examples of SB99 spectra 
(high and low--resolution) in the LW bands, overlaid with the 76 
ground state LW transitions. (Relative transition strengths are 
shown by the product of the oscillator strength $f_{\rm osc}$ and 
dissociation fraction $f_{\rm diss}$.) 
Because previous studies of the photodissociation rate have 
derived ${\rm \beta_{\rm approx}}$ from the low--resolution 
SB99 spectra, we first compare the accuracy of these approximations
using the low--resolution spectra in our full frequency--dependent
rate calculation. 

\begin{figure}
  \includegraphics[clip=true,trim=0.in 0.3in 0.3in 0.in,
    height=2.8in,width=3.2in]{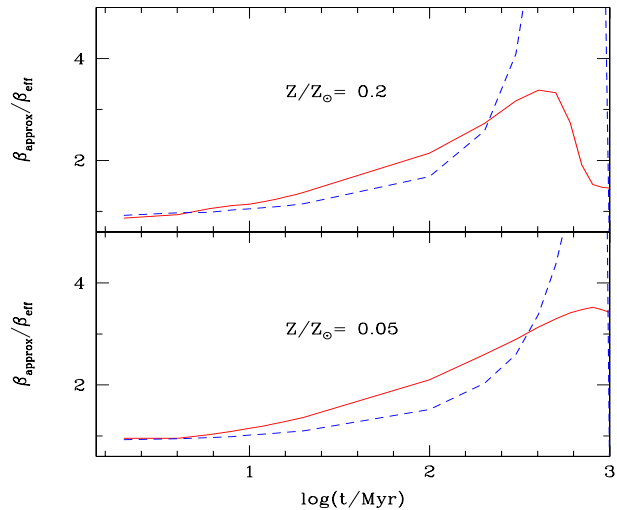}
  \caption{Ratio of the Lyman--Werner photodissociation rate,
    parameterised by the dimensionless parameter $\beta$
    (eq.~\ref{Eq:DefineBeta}), from an approximate method ($\beta_{\rm
      approx}$) to the full, frequency--dependent calculation
    ($\beta_{\rm eff}$). In all cases the spectrum is the
    low--resolution UV from a {\sc starburst99} instantaneous burst
    with metallicity as indicated on the figure.  The dissociation
    rate assuming either the average flux in the LW bands ($\beta_{\rm
      av}$) or a flat spectrum ($\beta_{\rm flat}$) with $J_{\rm LW} =
    J {\rm (12.4ev)}$ both overestimate the ``true'' rate by a factor
    of $\sim 2$ at 100 Myr after the burst and a larger factor for
    older bursts. This is due to the strong attenuation of the flux
    near the Lyman--limit (line blanketing), which is not captured by
    the approximations of flat spectra in the LW bands.}
  \label{Fig:CompareRates1}
\end{figure}

The results in Figure~\ref{Fig:CompareRates1} show that 
$\beta_{\rm flat}$ (solid red curve) and $\beta_{\rm avg}$ 
(dashed blue curves) are good approximations for the early 
(of order 10 Myr) SB99 spectra which are relatively flat in 
the LW bands, compared to later times. Both over--estimate 
the photodissociation rate by a factor of $\sim 2$ by 100 Myr 
and increasing thereafter, due to the significant dips in 
the low--resolution spectra near the Lyman--limit. 
We note that although the errors in these common methods of
normalizing are relatively small at early times, in the DCBH 
scenario, even factor of $\lsim 2$ errors in $k_{\rm LW}$ and
the corresponding critical flux leads to large errors in 
the predicted numbers of candidate halos \citep{Dijkstra+08}.

\begin{figure}
  \includegraphics[clip=true,trim=0.in 0.3in 0.3in 0.in,
    height=2.8in,width=3.2in]{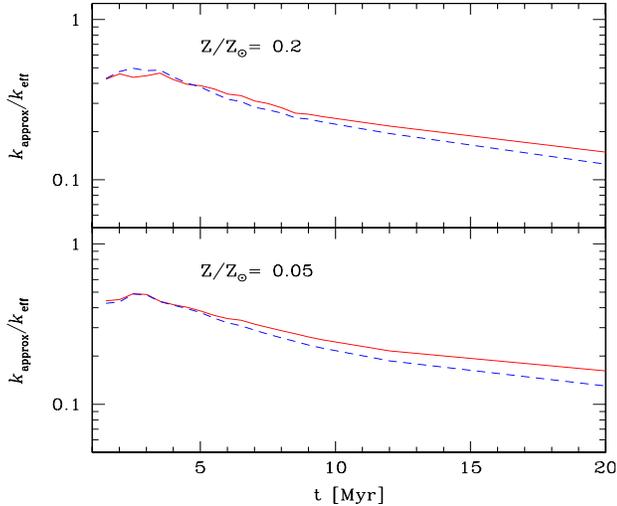}
  \caption{Ratio of the Lyman--Werner photodissociation rate from 
frequency--dependent calculation using the high--resolution UV 
{\sc starburst99} spectra ($k_{\rm eff}$) to the rate using the
low--resolution spectrum and normalizing assuming either (i)
a flat spectrum with 12.4eV normalization (red solid line) or 
(ii) average flux in the LW bands (blue dashed line).}
  \label{Fig:CompareRates2}
\end{figure}
Note that, as shown in Figure \ref{Fig:Spectra2}, beyond resolving 
individual lines, the high-resolution spectra tend to have higher 
mean flux levels than the low-resolution spectra. These two effects 
can lead to significant differences in the predicted photodissociation 
rates. In Figure \ref{Fig:CompareRates2} we show a comparison of the
(frequency--dependent) rates calculated using the high--resolution 
UV spectra (``$k_{\rm eff}$'') with those obtained from the approximate 
methods (flat or average) and low--resolution spectra (``$k_{\rm approx}$'').
This gives an indication of the overall error made in most previous 
studies, which use the latter. In this case, the effect of (modestly)
over--estimating $\beta$ is counterracted by the smaller flux levels
of the high--resolution spectra, and the upshot is the overall rate is
underestimated by a factor of a few. We include the comparison only up
to 20 Myr, after which the high--resolution UV spectra are no longer
reliable (C. Leitherer, private communication). 

\subsection{Critical Curve} 

\begin{figure}
  \includegraphics[clip=true,trim=0.in 0.1in 0.3in 0.in,
    height=2.8in,width=3.2in]{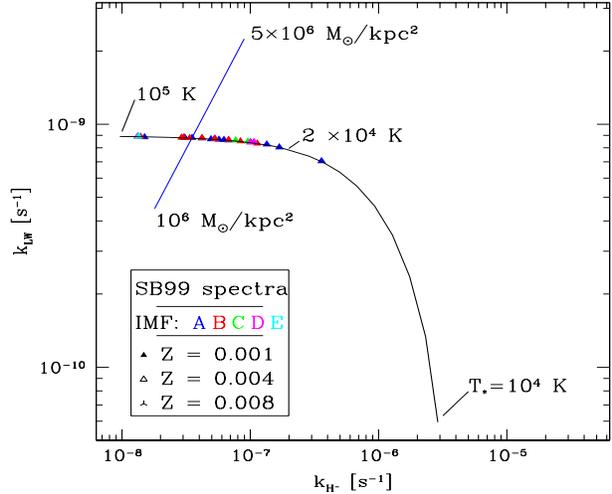}
  \caption{The ``critical curve'' (in black) shows the minimum combination of
    ${\rm H_2}$ photodissociation and ${\rm H^-}$--photodetachment
    rates required to prevent ${\rm H_2}$--cooling in an atomic
    cooling halo. 
    Above and to the right of the black curve is the region in
    which the halo remains hot enough to be a candidate for a direct
    collapse black hole. Several points are marked with the
    temperature of a blackbody spectrum required to produce the
    ratio of the two rates along the critical curve. Each triangle represents
    the rates produced by a particular {\sc starburst99} model
    spectrum with metallicity and IMF as indicated in the legend. 
    The (blue) diagonal line segment in the upper left shows the rates for SB99 models
    with varying stellar mass ($10^6 {\rm M_\odot}$ at lower left 
    and $5 \times 10^6 {\rm M_\odot}$ at top right), for IMF ``A''
    at 10 Myr after the burst and at a distance of 1 kpc. Depending 
    on the assumed efficiency of star formation, this indicates a 
    required (critical) halo mass of at least $10^{8} {\rm M_\odot}$ 
    for the illuminating galaxy.}
  \label{Fig:CriticalCurve}
\end{figure}

In general, ${\rm H_2}$ dissociation, and therefore the thermal
history, of gas in a primordial atomic cooling halo is controlled by
two rates: ${\rm H_2}$--photodissociation by LW photons ($k_{\rm LW}$)
and ${\rm H^-}$--photodetachment by IR photons ($k_{\rm H^-}$), since
${\rm H^-}$ is an intermediate in ${\rm H_2}$ formation. While a
threshold flux, $J_{\rm crit}$, has most often been used to determine
whether ${\rm H_2}$--cooling should be suppressed, this can be
misleading, and it requires the computation of $J_{\rm crit}$ specific
to each spectrum.  $J_{\rm crit}$ values have been estimated in the
literature for idealised spectra with fixed shapes, such as a
blackbody or power--law, but have then subsequently been adopted for
different spectral shapes, without re-computing the appropriate
$J_{\rm crit}$ values \citep[e.g.][]{Agarwal+12,DFM14}.  For
realistic, time--evolving spectra, such as those generated in
population synthesis models, both $k_{\rm LW}$ and $k_{\rm H^-}$ need
to be re-computed and specified to determine whether ${\rm
  H_2}$--cooling is suppressed. Although a corresponding $J_{\rm
  crit}$ can be computed, as well, this, as we have emphasized, is
unnecessary and misleading, and has led to erronous assumptions in
previous works.  We therefore propose to replace the ``critical flux''
with the spectrum-independent ``critical curve,'' as shown in Figure
\ref{Fig:CriticalCurve}.

The solid curve in Figure \ref{Fig:CriticalCurve} shows the minimum
combination of $k_{\rm LW}$ and $k_{\rm H^-}$ required to keep the gas
hot in our one--zone models. This curve results from the cooling
behavior of the gas in our one--zone models, which follows the
familiar bifurcated path: with a sufficiently strong radiation field,
the ${\rm H_2}$--fraction remains too low for cooling to occur, and
the gas is near the virial temperature of the halo throughout the
initial stages of collapse (up to ${\rm n \gsim 10^7~
  cm^{-3}}$). Below the threshold, the ${\rm H_2}$ fraction reaches
$\sim 10^{-3}$, cooling the gas to $\sim$ a few $\times 10^2$K.  On
the critical curve in Figure \ref{Fig:CriticalCurve}, for all points
above (rightward) of the curve, ${\rm H_2}$--cooling is suppressed,
while for those below (leftward) of the curve, ${\rm H_2}$--cooling
occurs and the gas temperature falls below $10^3$K. {\em The curve
  itself depends only on the details of the one--zone model, and does
  not depend on the shape of the irradiating spectrum.} 
The location where each individual model spectrum falls relative to 
the critical curve does, however, depend on the spectral shape.''

For ease of comparison with previous studies, we have marked
on the critical curve the positions of blackbodies of several 
different temperatures. It is worth noting that all blackbodies 
with temperatures $\gsim 20,000$K lie in the flat portion of 
the curve, i.e. where photodissociation is the dominant 
mechanism for suppressing ${\rm H_2}$--cooling, while 
${\rm H^-}$--photodetachment is only important for the coolest 
blackbodies considered, $\sim 10^4$K. While the T4 type has
often been used in previous studies to approximate Pop II, the
positions of the SB99 spectra on the critical curve underscore
that it is not a good representation of Pop II; further, the 
increased fraction of mass in stars remains a problem for 
achieving ${\rm H_2}$--cooling suppression from populations 
with the softest (T4 type) spectra \citep{WGH12}. 

The points in Figure \ref{Fig:CriticalCurve} are the results of 
our one--zone model using individual SB99 spectra for the illuminating 
radiation field, each normalised with ${\rm M_\ast / d^2}$ so as to 
coincide with the critical curve. SB99 bursts with all five IMFs and 
metallicities ${\rm Z/Z_\odot = 0.05 - 0.2}$ are shown. 
While the SB99 spectra at early times ($\lsim 10$Myr) and
low--metallicity (${\rm Z/Z_\odot = 0.05}$) are comparable 
on the critical curve to higher temperature blackbodies, 
indicating that LW dissociation dominates the suppression of 
the ${\rm H_2}$ abundance, the IR radiation becomes important 
before 20 Myr for models with modest metallicities 
(${\rm Z/Z_\odot = 0.2}$).

Our ``critical curve'' 
differs from that of \citet{Agarwal+16} due to our use of a
reduced self--shielding column density, (see \S~\ref{Sec:Model}),
as discussed above and in \citet{WHB11}. For the SB99 spectra, 
our results also differ due to our frequency--dependent calculation
of the photodissociation rate. Our $k_{\rm LW}$ is smaller in general, 
for a given model, than in other studies, since ${\beta_{\rm flat}}$ 
and ${\beta_{\rm avg}}$ overestimate the actual rate. 
Our results are broadly consistent with those of \citet{Sugimura+14}, 
although, as previously noted, we have chosen not to explicitly use 
$J_{\rm crit}$, as is done in their Figures 5 and 6. Again, the 
primary differences are due to our use of a reduced self--shielding 
column density, which yields a smaller $J_{\rm crit}$ for a given 
model; however, the overall trends in our results are broadly consistent. 

\begin{figure}
  \includegraphics[clip=true,trim=0.in 0.1in 0.3in 0.in,
    height=2.8in,width=3.2in]{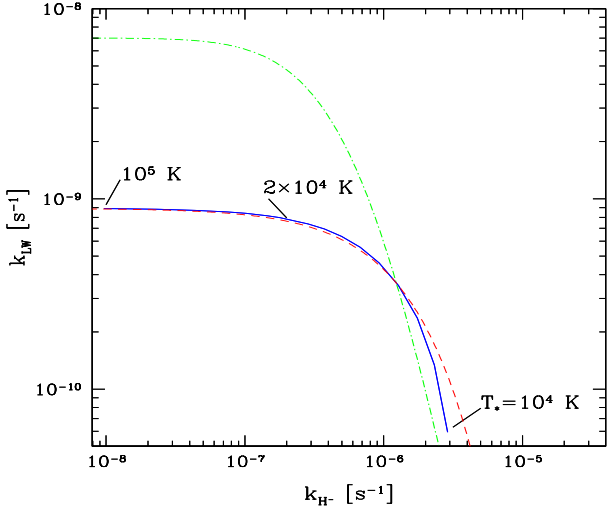}
  \caption{Our fitting formula (eq.~\ref{Eq:FittingFormula}) for the
    critical curve is shown in red (dashed), with the curve from our 
    one--zone calculation in blue (solid). A fitting formula from a 
    previous study \citep{Agarwal+16} is shown in green (dot--dash).}
  \label{Fig:FittingFormula}
\end{figure}

In Figure \ref{Fig:FittingFormula} we show our new fitting formula
(blue dashed curve) alongside the results of our one--zone modeling
(red solid curve) and the fit provided by \citet{Agarwal+16} (green
dot-dash curve). Our fit to the critical curve is given by:
\begin{equation} 
k_{\rm LW} = \frac{10^{\rm -A}}{(1 + k_{\rm H^-}/{\rm C})^{\rm D}}
\times {\rm exp}(-{\rm B} * k_{\rm H^-}/{\rm C}),
\label{Eq:FittingFormula} 
\end{equation}   
with A=9.05, B=0.95, C=$1.6 \times 10^{-6}$, and D=0.3. This fit 
is accurate to a few percent for $k_{\rm H^-} \lsim 2 \times 10^{-6}$,
corresponding to blackbody temperatures $\gsim 12,000$K
and the majority of SB99 results (which are leftward of that mark).
It underestimates $k_{\rm LW}$ by a factor of two at 
$k_{\rm H^-} = 3 \times 10^{-6}$ (${\rm T_{BB}} = 10^4$K). 

\subsection{Critical Star Formation in Illuminating Galaxy} 
\label{Sec:MCrit}
\begin{figure}
  \includegraphics[clip=true,trim=0.in 3.6in 0.3in 0.in,
    height=2.in,width=3.2in]{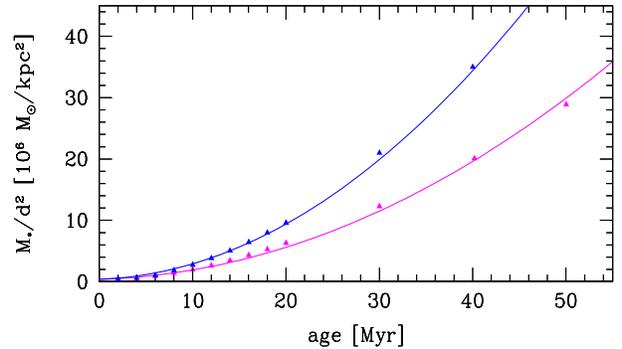}
  \caption{Our fitting formulae for 
$\Big({\rm \frac{M}{d^2}}\Big)_{\rm crit}$ are shown (solid curves) 
with the data (triangles) from our one--zone models. In both cases we 
used IMF A for the SB99 spectra, with Z = 0.001 (0.008) shown in 
magenta (blue).  The quadratic fit parameters are provided in 
\S~\ref{Sec:MCrit}.}
  \label{Fig:FittingFormula2}
\end{figure}

In order to easily utilize our results in semi-analytic models and
simulations of early structure formation, in
Table~\ref{tbl:CriticalHaloProperties} we provide the required stellar
mass in an illuminating galaxy with a given IMF and metallicity,
located at a distance $d$ from an atomic cooling halo, to prevent
${\rm H_2}$--cooling: $\Big({\rm \frac{M}{d^2}}\Big)_{\rm crit}$.
This allows identification of DCBH candidate halo pairs without
calculating individually the photodissociation and photodetachment
rates.
For two of our models (IMF A at Z = 0.001,0.008) we also found analytic 
fits for $\Big({\rm \frac{M}{d^2}}\Big)_{\rm crit}$, shown in Figure 
\ref{Fig:FittingFormula2}. These are quadratic fits with a=0.011 (0.02),
b=0.04 (0.05), c=0.4 (0.4)  for Z = 0.001 (0.008), with the usual 
convention used for the quadratic coefficients, ${\rm y = ax^2 + bx + c}$.
Both fitting functions are accurate to $<10$ per cent up to 50 Myr. 

\begin{table}
  \begin{center}
    \caption{The required ratio of stellar mass and squared-distance $\left[
        \frac{M_*/10^6 {\rm M_\odot}}{(d/{\rm kpc})^2}\right]_{\rm crit}$ for a
      starburst galaxy to prevent ${\rm H_2}$--cooling in a
      neighboring atomic cooling halo, obtained from the one--zone
      model.}
    \label{tbl:CriticalHaloProperties}
    \begin{tabular*}{0.475\textwidth}{@{\extracolsep{\fill}} l l l l l }
      \hline
      IMF & age (Myr) & Z = 0.001 & Z = 0.004 & Z = 0.008 \\
      \hline
      A  & 6  & 0.917& 0.994& 1.07 \\
         & 10 & 1.96 & 2.36 & 2.74 \\
         & 20 & 6.31 & 7.99 & 9.59 \\
      \hline
      B  & 6  & 0.316& 0.339& 0.359\\
         & 10 & 0.746& 0.893& 1.04 \\
         & 20 & 5.52 & 7.05 & 8.48 \\
      \hline
      C  & 6 &  0.84 & 1.27 & 3.12 \\
      \hline
      D  & 6 & 11.1  & 2.07 &  42.1 \\
      \hline
      E & 2 & 1.29  & 3.47 & 1.01 \\
      \hline\\
    \end{tabular*}
  \end{center}
\end{table}

\subsection{Importance of Level Populations}
\label{Sec:Caveats} 
The most significant simplification we made in this study is to assume
the ro-vibrational level populations of ${\rm H_2}$ are well described
by a Boltzmann distribution in the rotational levels (J = 0-29) of the
ground vibrational state (v=0). Deviations from LTE can significantly
affect the self--shielding behavior of ${\rm H_2}$ and therefore
change the critical curve shown in Figure \ref{Fig:CriticalCurve}. In
the early phases of gravitational collapse of an atomic cooling halo,
the ${\rm H_2}$ in v=0 would not have reached equilibrium, since the
critical densities are already ${\rm n \sim 10^3-10^4}~{\rm cm^{-3}}$
for the first few rotational levels.

At the relevant temperatures for the early collapse, i.e. a few
thousand Kelvin, high J states ($\rm J \gsim 5-10$) are negligibly
populated, so this does not introduce significant errors in the
optically--thin ${\rm H_2}$--photodissociation rate. As discussed in
\citet{WGH11}, however, the more levels are populated, the less
effective ${\rm H_2}$ self--shielding becomes, so the optically--thick
rate is more sensitive to the assumed distribution over ro-vibrational
states.  For example, in their study using the publicly available
package {\sc cloudy} to resolve level populations of ${\rm H_2}$,
\citet{RSO14} find differences compared to the optically--thick LTE
rate of $\sim$ a factor of a few for ${\rm T_{gas} = 5000}$K and ${\rm
  n_{tot} = 10^2~cm^{-3}}$ at very high column densities, ${\rm N_{H2}
  \gsim 10^{19}~cm^{-2}}$ with much smaller errors at the relevant
column densities for the early stages of collapse.

\section{Conclusions}
\label{Sec:Conclusions}
 
In most studies of the direct collapse black hole scenario, a single 
parameter $J_{\rm crit}$ has been used to delineate the threshold
flux above which ${\rm H_2}$--cooling is prevented in an atomic 
cooling--halo. 
However, this single-parameter assumption can be misleading, as the
value of $J_{\rm crit}$ depends on the spectral shape, and therefore
needs to be re-computed for each incident spectrum.  As long as the
spectrum is assumed to have a fixed shape (in time), such as a
blackbody or power--law, specifying the flux at a single frequency
(e.g. the Lyman--limit) fixes all relevant photo--chemical rates. In a
more general treatment, the ratio of the flux in the Lyman--Werner
bands to that at the energies relevant for ${\rm
  H^-}$--photodetachment ($\sim 1-2$eV), is dependent on the details
of the spectrum, which depend on the mass function, metallicity, and
age of the irradiating galaxy.  Although a new $J_{\rm crit}$ could be
computed for each case, we here emphasize that this is unnecessary,
and advocate the use, instead of a two-dimension critical curve
(Figure~\ref{Fig:CriticalCurve}).

Using the population synthesis package {\sc starburst99} to generate
realistic spectra, we investigated the relative importance of LW
photodissociation and ${\rm H^-}$--photodetachment in suppressing the
${\rm H_2}$ abundance and thereby preventing cooling below the virial
temperature of a halo. Very few studies have used spectra from
population synthesis modeling, and those that have
\citep{Sugimura+14,AK15,Agarwal+16} relied on simplifications to
derive a LW photodissociation rate, rather than performing a full,
frequency--dependent calculation. We show in \S~\ref{Sec:Results} that
these simplifications overestimate the true photodissociation rate by
a factor a few in the first $\sim 20$ Myr of a burst.

We provide a fitting formula to the critical curve from our one--zone
modeling (see Fig.~\ref{Fig:FittingFormula} and
Eq.~\ref{Eq:FittingFormula}), which includes important updates to the
photochemical model and is accurate at the percent level.  The
resulting critical stellar masses
(Table~\ref{tbl:CriticalHaloProperties} and Figure~\ref{Fig:FittingFormula2}) 
can be used directly in semi-analytic models and simulations of 
early structure formation.

\section*{Acknowledgments}

We thank S. Glover for helpful comments on an earlier version of this manuscript. We acknowledge support from NASA grants NNX15AB19G (to ZH) and NNX15AB20G (to GB), NSF grant 1312888 (to GB), the NSF Graduate Research Fellowship Program (JWG) and a Simons Fellowship for Theoretical Physics (ZH).

\bibliography{H2}

\begin{thebibliography}{}

\bibitem[\protect\citeauthoryear{{Abel}, {Anninos}, {Zhang} \& {Norman}}{{Abel}
  et~al.}{1997}]{Abel+97}
{Abel} T.,  {Anninos} P.,  {Zhang} Y.,    {Norman} M.~L.,  1997, \nat, 2, 181

\bibitem[\protect\citeauthoryear{{Abel}, {Bryan} \& {Norman}}{{Abel}
  et~al.}{2002}]{ABN02}
{Abel} T.,  {Bryan} G.~L.,    {Norman} M.~L.,  2002, Science, 295, 93

\bibitem[\protect\citeauthoryear{{Abgrall}, {Roueff} \& {Drira}}{{Abgrall}
  et~al.}{2000}]{ARD00}
{Abgrall} H.,  {Roueff} E.,    {Drira} I.,  2000, \aaps, 141, 297

\bibitem[\protect\citeauthoryear{{Abgrall}, {Roueff}, {Launay}, {Roncin} \&
  {Subtil}}{{Abgrall} et~al.}{1993}]{ARL93a}
{Abgrall} H.,  {Roueff} E.,  {Launay} F.,  {Roncin} J.~Y.,    {Subtil} J.~L.,
  1993, \aaps, 101, 273

\bibitem[\protect\citeauthoryear{{Agarwal}, {Dalla Vecchia}, {Johnson},
  {Khochfar} \& {Paardekooper}}{{Agarwal} et~al.}{2014}]{Agarwal+14}
{Agarwal} B.,  {Dalla Vecchia} C.,  {Johnson} J.~L.,  {Khochfar} S.,
  {Paardekooper} J.-P.,  2014, \mnras, 443, 648

\bibitem[\protect\citeauthoryear{{Agarwal} \& {Khochfar}}{{Agarwal} \&
  {Khochfar}}{2015}]{AK15}
{Agarwal} B.,  {Khochfar} S.,  2015, \mnras, 446, 160

\bibitem[\protect\citeauthoryear{{Agarwal}, {Khochfar}, {Johnson}, {Neistein},
  {Dalla Vecchia} \& {Livio}}{{Agarwal} et~al.}{2012}]{Agarwal+12}
{Agarwal} B.,  {Khochfar} S.,  {Johnson} J.~L.,  {Neistein} E.,  {Dalla
  Vecchia} C.,    {Livio} M.,  2012, \mnras, 425, 2854

\bibitem[\protect\citeauthoryear{{Agarwal}, {Smith}, {Glover}, {Natarajan} \&
  {Khochfar}}{{Agarwal} et~al.}{2016}]{Agarwal+16}
{Agarwal} B.,  {Smith} B.,  {Glover} S.,  {Natarajan} P.,    {Khochfar} S.,
  2016, \mnras, 459, 4209

\bibitem[\protect\citeauthoryear{{Ahn}, {Shapiro}, {Iliev}, {Mellema} \&
  {Pen}}{{Ahn} et~al.}{2009}]{Ahn+09}
{Ahn} K.,  {Shapiro} P.~R.,  {Iliev} I.~T.,  {Mellema} G.,    {Pen} U.,  2009,
  \apj, 695, 1430

\bibitem[\protect\citeauthoryear{{Begelman}, {Volonteri} \& {Rees}}{{Begelman}
  et~al.}{2006}]{BVR06}
{Begelman} M.~C.,  {Volonteri} M.,    {Rees} M.~J.,  2006, \mnras, 370, 289

\bibitem[\protect\citeauthoryear{{Bromm}, {Coppi} \& {Larson}}{{Bromm}
  et~al.}{2002}]{BCL02}
{Bromm} V.,  {Coppi} P.~S.,    {Larson} R.~B.,  2002, \apj, 564, 23

\bibitem[\protect\citeauthoryear{{Bromm} \& {Loeb}}{{Bromm} \&
  {Loeb}}{2003}]{BL03}
{Bromm} V.,  {Loeb} A.,  2003, \apj, 596, 34

\bibitem[\protect\citeauthoryear{{Dijkstra}, {Ferrara} \&
  {Mesinger}}{{Dijkstra} et~al.}{2014}]{DFM14}
{Dijkstra} M.,  {Ferrara} A.,    {Mesinger} A.,  2014, \mnras, 442, 2036

\bibitem[\protect\citeauthoryear{{Dijkstra}, {Haiman}, {Mesinger} \&
  {Wyithe}}{{Dijkstra} et~al.}{2008}]{Dijkstra+08}
{Dijkstra} M.,  {Haiman} Z.,  {Mesinger} A.,    {Wyithe} J.~S.~B.,  2008,
  \mnras, 391, 1961

\bibitem[\protect\citeauthoryear{{Field}, {Somerville} \& {Dressler}}{{Field}
  et~al.}{1966}]{FSD66}
{Field} G.~B.,  {Somerville} W.~B.,    {Dressler} K.,  1966, \araa, 4, 207

\bibitem[\protect\citeauthoryear{{Glover}}{{Glover}}{2015a}]{Glover15a}
{Glover} S.~C.~O.,  2015a, \mnras, 451, 2082

\bibitem[\protect\citeauthoryear{{Glover}}{{Glover}}{2015b}]{Glover15b}
{Glover} S.~C.~O.,  2015b, \mnras, 453, 2901

\bibitem[\protect\citeauthoryear{{Greif}, {Springel}, {White}, {Glover},
  {Clark}, {Smith}, {Klessen} \& {Bromm}}{{Greif} et~al.}{2011}]{Greif+11}
{Greif} T.~H.,  {Springel} V.,  {White} S.~D.~M.,  {Glover} S.~C.~O.,  {Clark}
  P.~C.,  {Smith} R.~J.,  {Klessen} R.~S.,    {Bromm} V.,  2011, \apj, 737, 75

\bibitem[\protect\citeauthoryear{{Habouzit}, {Volonteri}, {Latif}, {Dubois} \&
  {Peirani}}{{Habouzit} et~al.}{2016}]{Habouzit+16}
{Habouzit} M.,  {Volonteri} M.,  {Latif} M.,  {Dubois} Y.,    {Peirani} S.,
  2016, \mnras, 463, 529

\bibitem[\protect\citeauthoryear{{Haiman}}{{Haiman}}{2013}]{SMBHreview13}
{Haiman} Z.,  2013, in {Wiklind} T.,  {Mobasher} B.,   {Bromm} V.,  eds,
  Astrophysics and Space Science Library Vol.~396 of Astrophysics and Space
  Science Library, {The Formation of the First Massive Black Holes}.
p.~293

\bibitem[\protect\citeauthoryear{{Haiman}, {Abel} \& {Rees}}{{Haiman}
  et~al.}{2000}]{HAR00}
{Haiman} Z.,  {Abel} T.,    {Rees} M.~J.,  2000, \apj, 534, 11

\bibitem[\protect\citeauthoryear{{Haiman}, {Rees} \& {Loeb}}{{Haiman}
  et~al.}{1997}]{HRL97}
{Haiman} Z.,  {Rees} M.~J.,    {Loeb} A.,  1997, \apj, 476, 458

\bibitem[\protect\citeauthoryear{{Hosokawa}, {Hirano}, {Kuiper}, {Yorke},
  {Omukai} \& {Yoshida}}{{Hosokawa} et~al.}{2016}]{Hosokawa+15}
{Hosokawa} T.,  {Hirano} S.,  {Kuiper} R.,  {Yorke} H.~W.,  {Omukai} K.,
  {Yoshida} N.,  2016, \apj, 824, 119

\bibitem[\protect\citeauthoryear{{Hui} \& {Gnedin}}{{Hui} \&
  {Gnedin}}{1997}]{Hui+Gnedin}
{Hui} L.,  {Gnedin} N.~Y.,  1997, \mnras, 292, 27

\bibitem[\protect\citeauthoryear{{Hutchins}}{{Hutchins}}{1976}]{Hutchins76}
{Hutchins} J.~B.,  1976, \apj, 205, 103

\bibitem[\protect\citeauthoryear{{Inayoshi} \& {Haiman}}{{Inayoshi} \&
  {Haiman}}{2014}]{IH14}
{Inayoshi} K.,  {Haiman} Z.,  2014, \mnras, 445, 1549

\bibitem[\protect\citeauthoryear{{Inayoshi}, {Haiman} \& {Ostriker}}{{Inayoshi}
  et~al.}{2016}]{IHO15}
{Inayoshi} K.,  {Haiman} Z.,    {Ostriker} J.~P.,  2016, \mnras, 459, 3738

\bibitem[\protect\citeauthoryear{{Inayoshi} \& {Tanaka}}{{Inayoshi} \&
  {Tanaka}}{2015}]{IT15}
{Inayoshi} K.,  {Tanaka} T.~L.,  2015, \mnras, 450, 4350

\bibitem[\protect\citeauthoryear{{Koushiappas}, {Bullock} \&
  {Dekel}}{{Koushiappas} et~al.}{2004}]{KBD04}
{Koushiappas} S.~M.,  {Bullock} J.~S.,    {Dekel} A.,  2004, \mnras, 354, 292

\bibitem[\protect\citeauthoryear{{Kreckel}, {Bruhns}, {{\v C}{\'{\i}}{\v z}ek},
  {Glover}, {Miller}, {Urbain} \& {Savin}}{{Kreckel} et~al.}{2010}]{Kreckel+10}
{Kreckel} H.,  {Bruhns} H.,  {{\v C}{\'{\i}}{\v z}ek} M.,  {Glover} S.~C.~O.,
  {Miller} K.~A.,  {Urbain} X.,    {Savin} D.~W.,  2010, Science, 329, 69

\bibitem[\protect\citeauthoryear{{Latif}, {Schleicher}, {Bovino}, {Grassi} \&
  {Spaans}}{{Latif} et~al.}{2014}]{Latif+14}
{Latif} M.~A.,  {Schleicher} D.~R.~G.,  {Bovino} S.,  {Grassi} T.,    {Spaans}
  M.,  2014, \apj, 792, 78

\bibitem[\protect\citeauthoryear{{Leitherer}, {Schaerer}, {Goldader},
  {Delgado}, {Robert}, {Kune}, {de Mello}, {Devost} \& {Heckman}}{{Leitherer}
  et~al.}{1999}]{Leitherer+99}
{Leitherer} C.,  {Schaerer} D.,  {Goldader} J.~D.,  {Delgado} R.~M.~G.,
  {Robert} C.,  {Kune} D.~F.,  {de Mello} D.~F.,  {Devost} D.,    {Heckman}
  T.~M.,  1999, \apjs, 123, 3

\bibitem[\protect\citeauthoryear{{Lenzuni}, {Chernoff} \& {Salpeter}}{{Lenzuni}
  et~al.}{1991}]{Lenzuni+91}
{Lenzuni} P.,  {Chernoff} D.~F.,    {Salpeter} E.~E.,  1991, \apjs, 76, 759

\bibitem[\protect\citeauthoryear{{Lodato} \& {Natarajan}}{{Lodato} \&
  {Natarajan}}{2006}]{LN06}
{Lodato} G.,  {Natarajan} P.,  2006, \mnras, 371, 1813

\bibitem[\protect\citeauthoryear{{Martin}, {Schwarz} \& {Mandy}}{{Martin}
  et~al.}{1996}]{MSM96}
{Martin} P.~G.,  {Schwarz} D.~H.,    {Mandy} M.~E.,  1996, \apj, 461, 265

\bibitem[\protect\citeauthoryear{{Miyake}, {Stancil}, {Sadeghpour}, {Dalgarno},
  {McLaughlin} \& {Forrey}}{{Miyake} et~al.}{2010}]{Miyake+10}
{Miyake} S.,  {Stancil} P.~C.,  {Sadeghpour} H.~R.,  {Dalgarno} A.,
  {McLaughlin} B.~M.,    {Forrey} R.~C.,  2010, \apjl, 709, L168

\bibitem[\protect\citeauthoryear{{Mortlock}, {Warren}, {Venemans}, {Patel},
  {Hewett}, {McMahon}, {Simpson}, {Theuns}, {Gonz{\'a}les-Solares}, {Adamson},
  {Dye}, {Hambly}, {Hirst}, {Irwin}, {Kuiper}, {Lawrence} \&
  {R{\"o}ttgering}}{{Mortlock} et~al.}{2011}]{Mortlock+11}
{Mortlock} D.~J.,  {Warren} S.~J.,  {Venemans} B.~P.,  {Patel} M.,  {Hewett}
  P.~C.,  {McMahon} R.~G.,  {Simpson} C.,  {Theuns} T.,  {Gonz{\'a}les-Solares}
  E.~A.,  {Adamson} A.,  {Dye} S.,  {Hambly} N.~C.,  {Hirst} P.,  {Irwin}
  M.~J.,  {Kuiper} E.,  {Lawrence} A.,    {R{\"o}ttgering} H.~J.~A.,  2011,
  \nat, 474, 616

\bibitem[\protect\citeauthoryear{{Natarajan}}{{Natarajan}}{2014}]{NatarajanReview14}
{Natarajan} P.,  2014, General Relativity and Gravitation, 46, 1702

\bibitem[\protect\citeauthoryear{{Oh} \& {Haiman}}{{Oh} \&
  {Haiman}}{2002}]{OH02}
{Oh} S.~P.,  {Haiman} Z.,  2002, \apj, 569, 558

\bibitem[\protect\citeauthoryear{{Omukai}}{{Omukai}}{2001}]{O01}
{Omukai} K.,  2001, \apj, 546, 635

\bibitem[\protect\citeauthoryear{{Omukai}, {Schneider} \& {Haiman}}{{Omukai}
  et~al.}{2008}]{OSH08}
{Omukai} K.,  {Schneider} R.,    {Haiman} Z.,  2008, \apj, 686, 801

\bibitem[\protect\citeauthoryear{{Regan} \& {Haehnelt}}{{Regan} \&
  {Haehnelt}}{2009a}]{RH09a}
{Regan} J.~A.,  {Haehnelt} M.~G.,  2009a, \mnras, 396, 343

\bibitem[\protect\citeauthoryear{{Regan} \& {Haehnelt}}{{Regan} \&
  {Haehnelt}}{2009b}]{RH09b}
{Regan} J.~A.,  {Haehnelt} M.~G.,  2009b, \mnras, 393, 858

\bibitem[\protect\citeauthoryear{{Regan}, {Johansson} \& {Haehnelt}}{{Regan}
  et~al.}{2014}]{Regan+14}
{Regan} J.~A.,  {Johansson} P.~H.,    {Haehnelt} M.~G.,  2014, \mnras, 439,
  1160

\bibitem[\protect\citeauthoryear{{Regan}, {Johansson} \& {Wise}}{{Regan}
  et~al.}{2014}]{Regan+14b}
{Regan} J.~A.,  {Johansson} P.~H.,    {Wise} J.~H.,  2014, \apj, 795, 137

\bibitem[\protect\citeauthoryear{{Regan}, {Johansson} \& {Wise}}{{Regan}
  et~al.}{2016}]{Regan+16}
{Regan} J.~A.,  {Johansson} P.~H.,    {Wise} J.~H.,  2016, \mnras, 459, 3377

\bibitem[\protect\citeauthoryear{{Richings}, {Schaye} \&
  {Oppenheimer}}{{Richings} et~al.}{2014}]{RSO14}
{Richings} A.~J.,  {Schaye} J.,    {Oppenheimer} B.~D.,  2014, \mnras, 442,
  2780

\bibitem[\protect\citeauthoryear{{Saslaw} \& {Zipoy}}{{Saslaw} \&
  {Zipoy}}{1967}]{Saslaw+Zipoy}
{Saslaw} W.~C.,  {Zipoy} D.,  1967, \nat, 216, 976

\bibitem[\protect\citeauthoryear{{Schleicher}, {Spaans} \&
  {Glover}}{{Schleicher} et~al.}{2010}]{SSG10}
{Schleicher} D.~R.~G.,  {Spaans} M.,    {Glover} S.~C.~O.,  2010, \apjl, 712,
  L69

\bibitem[\protect\citeauthoryear{{Shang}, {Bryan} \& {Haiman}}{{Shang}
  et~al.}{2010}]{SBH10}
{Shang} C.,  {Bryan} G.~L.,    {Haiman} Z.,  2010, \mnras, 402, 1249

\bibitem[\protect\citeauthoryear{{Shapiro} \& {Kang}}{{Shapiro} \&
  {Kang}}{1987}]{SK87}
{Shapiro} P.~R.,  {Kang} H.,  1987, \apj, 318, 32

\bibitem[\protect\citeauthoryear{{Spaans} \& {Silk}}{{Spaans} \&
  {Silk}}{2006}]{SS06}
{Spaans} M.,  {Silk} J.,  2006, \apj, 652, 902

\bibitem[\protect\citeauthoryear{{Stecher} \& {Williams}}{{Stecher} \&
  {Williams}}{1967}]{SW67}
{Stecher} T.~P.,  {Williams} D.~A.,  1967, \apjl, 149, L29+

\bibitem[\protect\citeauthoryear{{Stenrup}, {Larson} \& {Elander}}{{Stenrup}
  et~al.}{2009}]{Stenrup+09}
{Stenrup} M.,  {Larson} {\AA}.,    {Elander} N.,  2009, \pra, 79, 012713

\bibitem[\protect\citeauthoryear{{Sugimura}, {Omukai} \& {Inoue}}{{Sugimura}
  et~al.}{2014}]{Sugimura+14}
{Sugimura} K.,  {Omukai} K.,    {Inoue} A.~K.,  2014, \mnras, 445, 544

\bibitem[\protect\citeauthoryear{{Visbal}, {Haiman} \& {Bryan}}{{Visbal}
  et~al.}{2014}]{VHB14b}
{Visbal} E.,  {Haiman} Z.,    {Bryan} G.~L.,  2014, \mnras, 445, 1056

\bibitem[\protect\citeauthoryear{{Volonteri} \& {Bellovary}}{{Volonteri} \&
  {Bellovary}}{2012}]{VBreview12}
{Volonteri} M.,  {Bellovary} J.,  2012, Reports on Progress in Physics, 75,
  124901

\bibitem[\protect\citeauthoryear{{Volonteri}, {Lodato} \&
  {Natarajan}}{{Volonteri} et~al.}{2008}]{VLN08}
{Volonteri} M.,  {Lodato} G.,    {Natarajan} P.,  2008, \mnras, 383, 1079

\bibitem[\protect\citeauthoryear{{Wise} \& {Abel}}{{Wise} \&
  {Abel}}{2008}]{WA08}
{Wise} J.~H.,  {Abel} T.,  2008, \apj, 685, 40

\bibitem[\protect\citeauthoryear{{Wishart}}{{Wishart}}{1979}]{Wishart79}
{Wishart} A.~W.,  1979, \mnras, 187, 59P

\bibitem[\protect\citeauthoryear{{Wolcott-Green} \& {Haiman}}{{Wolcott-Green}
  \& {Haiman}}{2011}]{WGH11}
{Wolcott-Green} J.,  {Haiman} Z.,  2011, \mnras, 412, 2603

\bibitem[\protect\citeauthoryear{{Wolcott-Green} \& {Haiman}}{{Wolcott-Green}
  \& {Haiman}}{2012}]{WGH12}
{Wolcott-Green} J.,  {Haiman} Z.,  2012, \mnras, 425, L51

\bibitem[\protect\citeauthoryear{{Wolcott-Green}, {Haiman} \&
  {Bryan}}{{Wolcott-Green} et~al.}{2011}]{WHB11}
{Wolcott-Green} J.,  {Haiman} Z.,    {Bryan} G.~L.,  2011, \mnras, 418, 838

\bibitem[\protect\citeauthoryear{{Yoshida}, {Abel}, {Hernquist} \&
  {Sugiyama}}{{Yoshida} et~al.}{2003}]{Yoshida+03}
{Yoshida} N.,  {Abel} T.,  {Hernquist} L.,    {Sugiyama} N.,  2003, \apj, 592,
  645

\end{thebibliography}

\label{lastpage}

\end{document}